\documentclass[runningheads]{llncs}
\usepackage{graphicx}
\usepackage{comment}
\usepackage{amsmath,amssymb} 
\usepackage{color}

\usepackage{subcaption}

\newcommand{\eg}{\textit{e.g.},}
\newcommand{\ie}{\textit{i.e.},}

\begin{document}
\pagestyle{headings}
\mainmatter
\def\ECCVSubNumber{1849}  

\title{JSSR: A Joint Synthesis, Segmentation, and Registration System for 3D Multi-Modal Image Alignment of Large-scale Pathological CT Scans} 

\titlerunning{JSSR System for 3D Multi-Modal Image Alignment}
%
\author{Fengze Liu\inst{1,2} \and
Jinzheng Cai\inst{1} \and   Yuankai Huo\inst{1,3}  \and  Chi-Tung Cheng\inst{4} \and  Ashwin Raju\inst{1} \and Dakai Jin\inst{1} \and  Jing Xiao\inst{5} \and Alan Yuille\inst{2} \and  Le Lu\inst{1} \and  ChienHung Liao\inst{4} \and Adam P. Harrison\inst{1} }
\authorrunning{F. Liu et al.}
%
\institute{PAII Inc., Bethesda MD 20817, USA \and 
Johns Hopkins University, Baltimore MD 21218, USA \and
Vanderbilt University, Nashville TN 37235, USA \and Chang Gung Memorial Hospital, Linkou, Taiwan, ROC \and Ping An Technology, Shenzhen, China}
\maketitle

\begin{abstract}
Multi-modal image registration is a challenging  problem that is also an important clinical task for many real applications and scenarios. As a first step in analysis, deformable registration among different image modalities is often required in order to provide complementary visual information. During registration, semantic information is key to match homologous points and pixels. Nevertheless, many conventional registration methods are incapable in capturing high-level semantic anatomical dense correspondences. In this work, we propose a novel multi-task learning system, JSSR, based on an end-to-end 3D convolutional neural network that is composed of a generator, a registration and a segmentation component. The system is optimized to satisfy the implicit constraints between different tasks in an unsupervised manner. It first synthesizes the source domain images into the target domain, then an intra-modal registration is applied on the synthesized images and target images. The segmentation module are then applied on the synthesized and target images, providing additional cues based on semantic correspondences. The supervision from another fully-annotated dataset is used to regularize the segmentation. We extensively evaluate JSSR on a large-scale medical image dataset containing 1,485 patient CT imaging studies of four different contrast phases (i.e., 5,940 3D CT scans with pathological livers) on the registration, segmentation and synthesis tasks. The performance is improved after joint training on the registration and segmentation tasks by $0.9\%$ and $1.9\%$ respectively compared to a highly competitive and accurate deep learning baseline. The registration also consistently outperforms conventional state-of-the-art multi-modal registration methods.
\end{abstract}

\section{Introduction}

Image registration attempts to discover a spatial transformation between a pair of images that registers the points in one of the images to the homologous points in the other image \cite{AOODS2009495}. Within medical imaging, registration often focuses on inter-patient/inter-study mono-modal alignment. Another important and (if not more) frequent focal point is multi-channel imaging, \eg{} dynamic-contrast computed tomography (CT), multi-parametric magnetic resonance imaging (MRI), or positron emission tomography (PET) combined with CT/MRI. In this setting, the needs of intra-patient multi-modal registration are paramount, given the unavoidable patient movements or displacements between subsequent imaging scans. For scenarios where deformable misalignments are present, \eg{} the abdomen, correspondences can be highly complex. Because different modalities provide complementary visual/diagnosis information, proper and precise anatomical alignment benefits human reader's radiological observation and is crucial for any downstream computerized analyses. However, finding correspondences between homologous points is usually not trivial because of the complex appearance changes across modalities, which may be conditioned on anatomy, pathology, or other complicated interactions.

Unfortunately, multi-modal registration remains a challenging task, particularly since ground-truth deformations are hard or impossible to obtain. Methods must instead learn transformations or losses that allow for easier correspondences between images. Unsupervised registration methods, like \cite{balakrishnan2019VoxelMorph,heinrich2012globally}, often use a local modality invariant feature to measure similarity. However these low-level features may not be universally applicable and cannot always capture high level semantic information. Other approaches use generative models to reduce the domain shift between modalities, and then apply registration based on direct intensity similarity \cite{tanner2018generative}. A different strategy learns registrations that maximize the overlap in segmentation labels \cite{balakrishnan2019VoxelMorph,hu2018weakly}. This latter approach is promising, as it treats the registration process similarly to a segmentation task, aligning images based on their semantic category. Yet, these approaches rely on having supervised segmentation labels in the first place for every deployment scenario. 

\begin{figure}[!t]
\begin{center}
    \includegraphics[width=\textwidth]{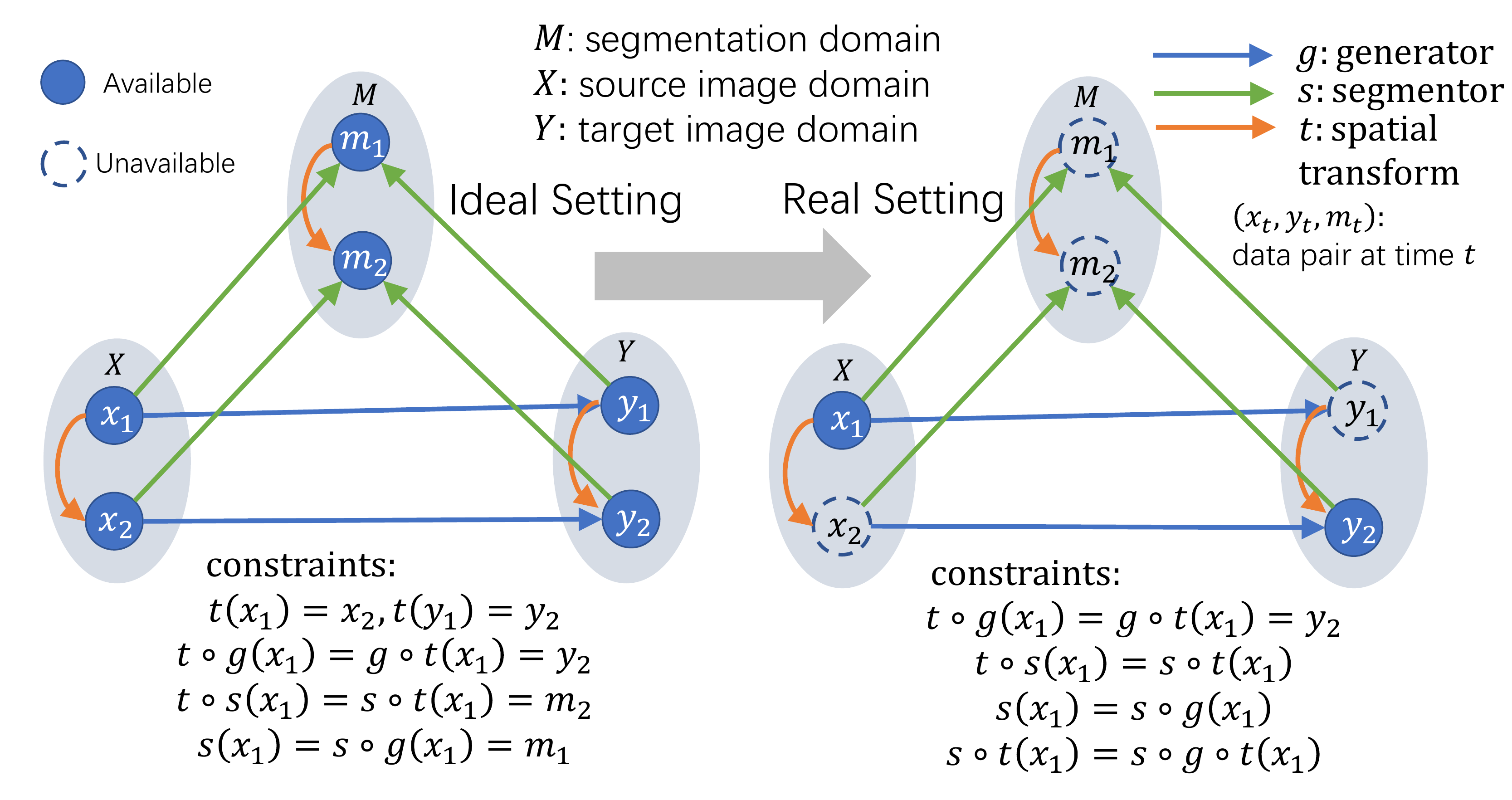}
\end{center}
\caption{
   The relationship between the synthesis, segmentation and registration tasks. In the ideal setting, spatially transformed examples from each domain, and their segmentation labels, are fully available. In more realistic settings, only one example is available from each domain, each under a different spatial transform. Moreover, segmentation labels are not available. Should segmentation, synthesis, and spatial transform \emph{mappings} be available, the constraints in the ideal case can be mapped to analogous constraints in the real case.
}
\label{motivation}
\end{figure}

Both the synthesis and segmentation approaches are promising, but they are each limited when used alone, especially when fully-supervised training data is not available, \ie{} no paired multi-modal images and segmentation labels, respectively. As Fig. \ref{motivation} elaborates, the synthesis, segmentation, and registration tasks are linked together and define implicit constraints between each other. That motivates us to develop a joint synthesis, segmentation, and registration (JSSR) system which satisfies these implicit constraints. JSSR is composed of a generator, a segmentation, and a registration component that performs all three tasks simultaneously. Given a fixed image and moving image from different modalities for registration, the generator can synthesize the moving image to the same modality of the fixed image, conditioned on the fixed image to better reduce the domain gap. Then the registration component accepts the synthesized image from the generator and the fixed image to estimate a deformation field. Lastly, the segmentation module estimates the segmentation map for the moving image, fixed image and synthesized image. During the training procedure, we optimize several consistency losses including (1) the similarity between the fixed image and the warped synthesized image; (2) the similarity between the segmentation maps of the warped moving image and the fixed image; (3) an  adversarial loss for generating high fidelity images; and (4) a smoothness loss to regularize the deformation field. To stop the segmentation module from providing meaningless segmentation maps, we regularize the segmentation by training it on fully supervised data obtained from a different source than the target data, e.g., public data. We evaluate our system on a large-scale clinical liver CT image dataset containing four phases per patient, for unpaired image synthesis, multi-modal image registration, and multi-modal image segmentation tasks. Our system outperforms the state-of-the-art conventional multi-modal registration methods and significantly improves the baseline model we used for the fother two tasks, validating the effectiveness of joint learning. 



We summarize our main contributions as follows:
\begin{itemize}
    \item We propose a novel joint learning approach for multi-modal image registration that incorporates the tasks of synthesis, registration and segmentation. Each task connects to the other two tasks during training, providing mutually reinforcing supervisory signals.
    \item We evaluate and validate the performance improvement of baseline methods for synthesis and segmentation after joint training by our system, demonstrating the effectiveness of our joint training setup and revealing the possibility of obtaining a better overall system by building upon and enhancing the baseline models.
    \item Our system consistently and significantly outperforms state-of-the-art conventional multi-modal registration approaches based on a large-scale multi-phase CT imaging dataset of 1,485 patients (each patient under four different intravenous contrast phases, i.e., 5,940 3D CT scans with various liver tumors). 
    \item While we use supervised data from \emph{single-phase} public data to regularize our segmentation, our method does not use or rely upon any manual segmentation labels from the target \emph{multi-phase} target CT imaging dataset. Compared to approaches expecting target segmentation labels, JSSR enjoys better scalability and generalizability  for varied clinical applications.
\end{itemize}

\section{Related Work}
\paragraph{Multi-modal Image Registration}
Multi-modal image registration has been widely studied and applied in medical imaging. Existing registration methods can be based on additional information, e.g., landmarks~\cite{landmark2001,landmark2019} or a surface~\cite{Yang2011}, or they can operate directly on voxel intensity values without any additional constraints introduced by the user or segmentation~\cite{maintz1996overview}. For voxel-based methods, there are two typical strategies. One is to transform each image using self-similarity measurements that are invariant across modalities. These include local self-similarities~\cite{shechtman2007matching} or the  
modality independent neighbourhood descriptor~\cite{heinrich2012mind}. Notably the DEEDS algorithm~\cite{heinrich2012globally,heinrich2013mrf,heinrich2015multi} employed a discrete dense displacement sampling for deformable registration using self-similarity context (SSC)~\cite{heinrich2013towards}. The other common strategy is to map both modalities into a shared space and measure the mono-modal difference. Prominent examples include mutual information \cite{maes1997multimodality} and normalized mutual information \cite{studholme1999overlap} similarity measures that can be applied directly on cross-modal images. However, such methods can suffer from low convergence rates and loss of spatial information. \cite{blendowski2019learning} employed a convolutional neural network (CNN) to learn modality invariant features using a small amount of supervision data. \cite{cao2017dual} used Haar-like features from paired multi-modality images to fit a random forest regression model for bi-directional image synthesis, and \cite{tanner2018generative,mahapatra2018deformable} applied CycleGANs to reduce the gap between modalities for better alignment. Recently \cite{arar2020unsupervised} developed a joint synthesis and registration framework on natural 2D images.


Recently a variety of deep learning-based registration methods have been proposed. Because  ground truth deformation fields are hard to obtain, unsupervised methods, like~\cite{de2017end,li2017non,de2019deep,balakrishnan2019VoxelMorph}, are popular. These all rely on a CNN with a spatial transformation function~\cite{jaderberg2015spatial}. These unsupervised methods mainly focus on mono-modal image registration. Some methods make use of correspondences between labelled anatomical structures to help the registration process~\cite{hu2018weakly}. \cite{balakrishnan2019VoxelMorph} also showed how the segmentation map can help registration. However, in many cases the segmentation map is not available, which motivates us to combine the registration and segmentation components together. 

\paragraph{Multi-task Learning Methods}
As the registration, synthesis, segmentation tasks are all related with each other, there are already several works that explore combining them together. \cite{tanner2018generative,mahapatra2018deformable,wei2019synthesis} used CycleGANs to synthesize multi-modal images into one modality, allowing the application of mono-modal registration methods. \cite{ketcha2019learning} projected multi-modal images into a shared feature space and registered based on the features. \cite{qin2019unsupervised} made use of a generative model to disentangle the appearance space from the shape space. \cite{li2019hybrid,xu2019deepatlas,qin2018joint} combined a segmentation model with a registration model to let them benefit each other, but the focus was on mono-modal registration. \cite{zhou2019hyper} performed supervised multi-phase segmentation based on paired multi-phase images but did not jointly train the registration and segmentation. \cite{zhang2018translating,Zheng2018PhaseCN,huo2018synseg} used a generative model to help guide the segmentation model. In contrast, our work combines \emph{all three} of the tasks together to tackle multi-modal registration problem in the most general setting where the deformation ground truth, paired multi-modal images and segmentation maps are \emph{all unavailable}.

\begin{figure}[!t]
\begin{center}
    \includegraphics[width=\textwidth]{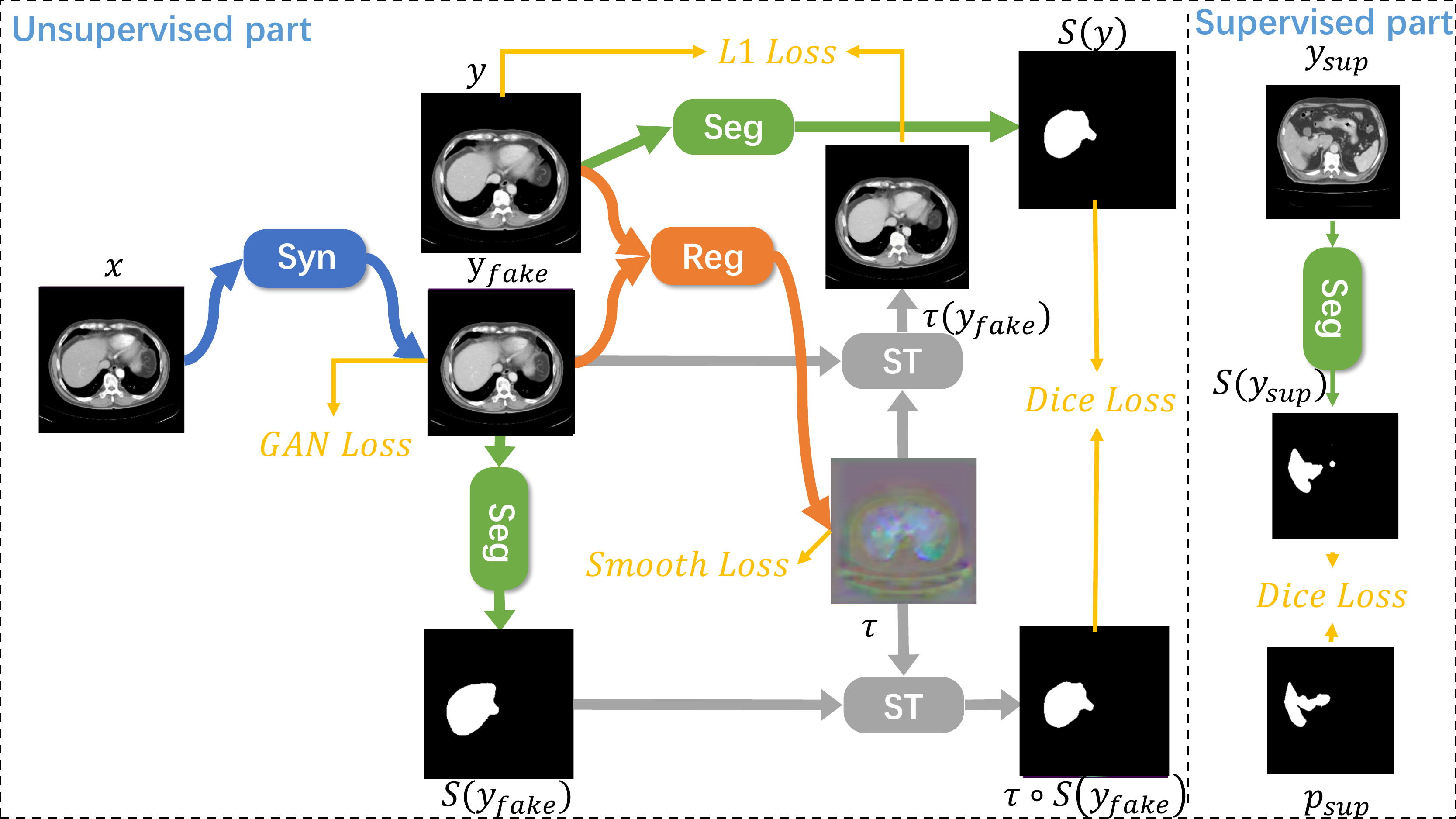}
\end{center}
\caption{
   The JSSR system. We denote the generator, segmentation, registration module and spatial transform as Syn, Seg, Reg and ST respectively. 
}
\label{main}
\end{figure}
\begin{figure}[!t]
\begin{center}
    \includegraphics[width=\textwidth]{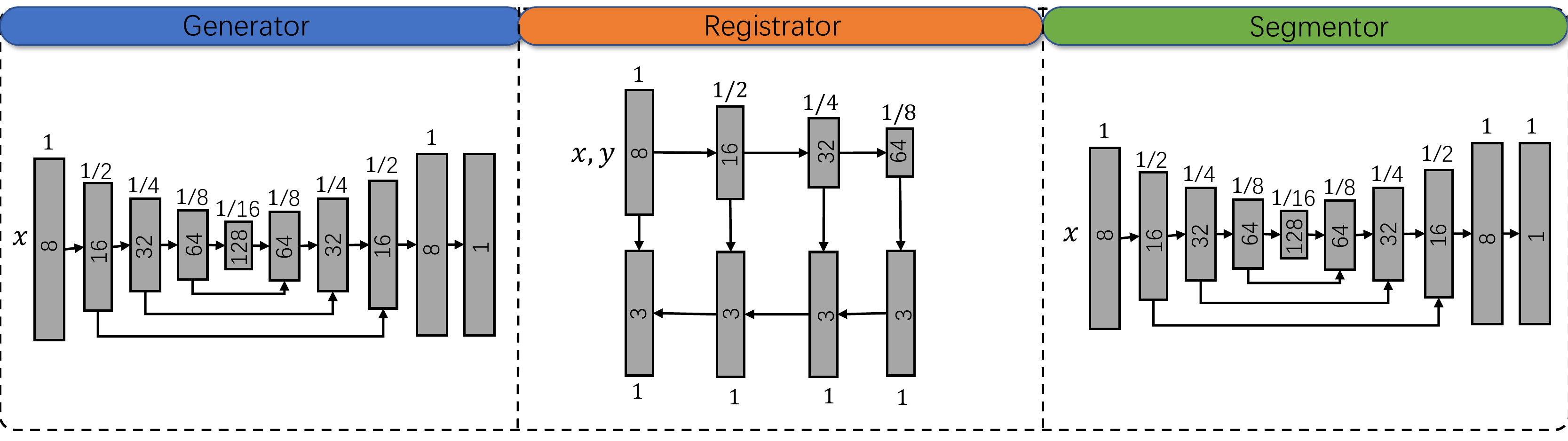}
\end{center}
\caption{
   The model structure for each component. We use a 3D PHNN~\cite{phnn} for registration and 3D VNet~\cite{milletari2016v} for segmentation and the generator.
}
\label{model}
\end{figure}
\section{Methodology}

Given a moving image $x\in\mathcal{X}$ and fixed image $y\in \mathcal{Y}$ from different modalities, \emph{but from the same patient}, we aim to find a spatial transformation function $\tau$ that corrects for any misalignments between the two. We tackle this multi-modal image registration problem in a fully unsupervised way to meet common applications settings, where none of the ground truth deformation fields, segmentation maps, or paired multi-modal images are available. As Fig.~\ref{motivation} depicts, image synthesis, segmentation and registration can be related together via a set of constraints. Motivated by this, we develop a system consisting of three parts: a generator $G$, a registration module $\Phi$ and a segmentation module $S$. By satisfying the constraints in Fig.~\ref{motivation}, we can satisfy the conditions for a correct registration, segmentation and image synthesis. During optimization, these three tasks will benefit from each other. Refer to Fig.~\ref{main} for the overall framework of our system. 

\subsection{Unpaired Image Synthesis}
\label{sec:synthesis}


Although good unpaired image synthesis works exist, e.g., \cite{huang2018multimodal}, they may generate a variety of different target domain images based on the random sampling. However, for registration, the synthesized images should have identical anatomical and other pertinent modality-invariant properties. Thus, a conditional synthesis is a natural choice.  Similar to \cite{isola2017image}, but without random noise, we use a GAN with a dual-input generative model $G$ which learns a mapping from $x,y$ to $\tau^{-1}(y)$, $G:\{x,y\}\xrightarrow{}\tau^{-1}(y)$. Here $\tau$ is the true deformation from $x$ to $y$, meaning the generator attempts to generate a version of $x$ that looks like $y$, but removing any spatial transformation between the two. In reality, $\tau$ itself must be estimated, which we will outline in Sec.~\ref{sec:mm_registration}. A discriminator $D$ is also equipped to detect the fake images from the generator.

The objective of the conditional GAN is
\begin{equation}
    \mathcal{L}_{GAN}(G,D)=E_y\log{D(y)}-E_{x,y}\log{D(G(x,y))}.
\end{equation}
In a classical paired GAN setup, we would use $E_y\log{D(\tau^{-1}(y))}$, but this is not available, so use unpaired synthesis, based on the assumption that spatial transform $\tau$ does not alter the likelihood of any one sample. We also add another appearance-based loss to benefit the GAN objective:
\begin{equation}
    \mathcal{L}^{syn}_{L1}(G)=E_{x,y}||\tau^{-1}(y)-G(x,y)||_1 .
\end{equation}
The final objective for the synthesis part is
\begin{equation}
    G^* = \arg \min_G\max_D \mathcal{L}^{syn}_{L1}(G)+\lambda_{syn} \mathcal{L}_{GAN}(G,D).
\end{equation}

\subsection{Multi-Modal Image Registration}
\label{sec:mm_registration}
For two images $x$ and $y$, the registration module learns a function $\Phi:{x,y}\xrightarrow{}\tau$ where $\tau$ is a spatial transformation function~\cite{jaderberg2015spatial}, also called the deformation field. For mono-modal registration, the $L_1$ loss can be used to estimate a deformation field that directly matches the intensities between the fixed image and warped image. Here we are registering two images from different modalities. \cite{balakrishnan2019VoxelMorph} proposed to use a cross-modal similarity measure like cross-correlation~\cite{avants2008symmetric}. Instead, if we assume a generative model is available to transform $x$ into the $\mathcal{Y}$ domain, then we can use a simple mono-modal similarity measure:
\begin{equation}
    \mathcal{L}^{reg}_{L1}(\Phi)=E_{x,y}||\tau(G(x,y))-y||_1,
\end{equation}
where $\tau=\Phi(G(x,y),y)$, and $G$ is the generator that synthesizes images from $\mathcal{X}$ to $\mathcal{Y}$. Another smoothness term is added to prevent non-realistic deformation:
\begin{equation}
    \mathcal{L}_{smooth}(\Phi)=E_{x,y}\sum_{v\in\Omega}||\nabla\tau_v||^2,
\end{equation}
where $v$ represents the voxel location and $\nabla\tau_v$ calculates the differences between neighboring voxels of $v$. We use the same implementation for the smoothness term as in \cite{balakrishnan2019VoxelMorph}. The final objective is:
\begin{equation}
    \Phi^* =  \arg \min_\Phi\mathcal{L}^{reg}_{L1}(\Phi) +\lambda_{reg}\mathcal{L}_{smooth}(\Phi)  \textrm{.}
\end{equation}

Of course, we cannot optimize this objective without a $G$. However, to get a good $G$, we need a good $\Phi$ as discussed in Sec.~\ref{sec:synthesis}, which makes this problem a chicken-and-egg conundrum. One way is to optimize the two objectives from the synthesis and registration modules together, which leads to

\begin{align}\label{bi}
\begin{split}
    \Phi^*,G^* &=\arg \min_{\Phi,G}\mathbb{F}(\Phi,G) \\
    & =\arg \min_{\Phi,G}\max_D\mathcal{L}^{reg}_{L1}(\Phi,G)+\mathcal{L}^{syn}_{L1}(\Phi,G) \\
    &+\lambda_{reg}\mathcal{L}_{smooth}(\Phi,G) +\lambda_{syn}\mathcal{L}_{GAN}(G,D) \\
    & \approx \arg \min_{\Phi,G}\max_D\  2\mathcal{L}^{reg}_{L1}(\Phi,G)\\
    &+\lambda_{reg}\mathcal{L}_{smooth}(\Phi,G) +\lambda_{syn}\mathcal{L}_{GAN}(G,D).
\end{split}
\end{align}


However, there is no guarantee that we can get the optimal solution by minimizing $\mathbb{F}(\Phi,G)$. Actually there is a trivial solution that minimizes $\mathbb{F}(\Phi,G)$, which is when $G(x,y)=y$ and $\Phi(G(x,y),y)=\Phi(y,y)=I$, i.e., the identity transform. To mitigate this, we add skip connections from the source domain to keep the spatial information in the structure of generator, as shown in Fig.~\ref{model}. 

\subsection{Multi-Modal Image Segmentation}
We enforce segmentation-based constraints for two reasons. Firstly, as noted in \cite{balakrishnan2019VoxelMorph}, the additional information of segmentation maps can help guide the registration process. However, \cite{balakrishnan2019VoxelMorph} assumes the segmentation maps are available for the target dataset, which we do not assume. Secondly, as noted by others~\cite{li2019hybrid,xu2019deepatlas,qin2018joint,zhang2018translating,Zheng2018PhaseCN}, synthesis and registration can benefit segmentation, which can help develop better segmentation models on datasets without annotation. 


We denote the segmentation model as a function $S:x\xrightarrow{}p$, where $p\in\mathcal{P}$ represents the segmentation map domain. Based on the constraint between synthesis, registration and segmentation tasks, we define the objective as: 
\begin{equation}\label{segunsup}
    \mathcal{L}_{dice}^{reg}(S,\Phi,G) =  E_{x,y}1-Dice[\tau(S(G(x,y))),S(y)],
\end{equation}
where $\tau=\Phi(G(x,y),y)$ and $Dice(x,y)=\frac{2x^T y}{x^T x+y^T y}$ is the widely used measurement for the similarity between two binary volumes.  This loss term connects three components together and in the experiments afterwards we show this crucial toward the whole system's performance. 

To make \eqref{segunsup} work properly, we need the segmentation to be as accurate as possible. However only with the consistency loss, the segmentation module is not able to learn meaningful semantic information. For instance, a segmentation module that predicts all background can trivially minimize \eqref{segunsup}. To avoid this, we use fully supervised data, e.g., from public sources, to regularize the segmentation. Importantly, because \eqref{segunsup} is only applied on the $\mathcal{Y}$ domain, we need only use supervised data from one modality, e.g., if we are registering dynamic contrast CT data, we need only fully-supervised segmentation maps from the more ubiquitous venous-phase CTs found in public data. Thus, the supervision loss is defined as
\begin{equation}
    \mathcal{L}_{dice}^{sup}(S) =  E_{y_{sup}}1-Dice[S(y_{sup}),p_{sup})],
\end{equation}
where $y_{sup}\in\mathcal{Y}$ is in the same modality with $y\in\mathcal{Y}$, but the two datasets do not overlap. $p_{sup}\in\mathcal{P}_{sup}$ is the corresponding annotation. The total loss provided by the segmentation module is
\begin{equation}
    \mathbb{H}(S,\Phi,G)=\mathcal{L}_{dice}^{reg}(S,\Phi,G) + \mathcal{L}_{dice}^{sup}(S) \textrm{.}
\end{equation}

\subsection{Joint Optimization Strategy}
Based on previous sections, the final objective for our whole system is
\begin{align}\label{tri}
\begin{split}
        \Phi^*,G^*,S^*=\arg\min_{\Phi,G,S}\mathbb{F}(\Phi,G)+\lambda_{seg}\mathbb{H}(S,\Phi,G).
    \end{split}
\end{align}

In order to provide all the components with a good initial point, we first train $S$ on the fully-supervised data, $\{y_{sup},p_{sup}\}$ and also train $\Phi$ and $G$ using \eqref{bi} on the unsupervised data. Finally, we jointly optimize all modules by \eqref{tri}. When optimizing \eqref{bi} and \eqref{tri}, we use the classic alternating strategy for  training GAN models, which alternately fixes $\Phi,G,S$ and optimizes for $D$ and then fixes $D$ and optimizes for the others.

\section{Experiments}
{\bf Datasets.} We conduct our main experiments on a large-scale dataset of 3D dynamic contrast multi-phase liver CT scans, 
extracted from the archives of the Chang Gung Memorial Hospital (CGMH) in Taiwan. The dataset is composed of 1485 patient studies and each studies consists of CT volumes of four different intravenous contrast phases: venous, arterial, delay, and non-contrast. The studied  population is composed of patients with liver tumors who underwent CT imaging examinations prior to an interventional biopsy, liver resection, or liver transplant. Our end goal is to develop a computer-aided diagnosis system to identify the pathological subtype of any given liver tumor. Whether the analysis is conducted by human readers or computers, all phases need to be precisely pre-registered to facilitate downstream analysis, which will observe  the dynamic contrast changes within liver tumor tissues across the sequential order of non-contrast, arterial, venous and delay CTs.   

The different phases are obtained from the CT scanner at different time points after the contrast media injection and will display different information according to the distribution of contrast media in the human body. The intensity value of each voxel in the CT image, measured by the Hounsfield Unit (HU), is an integer ranging from $-1000$HU to $1000$HU, which will also be affected by the density of contrast media. The volume size of the CT image is $512\times512\times L$, where $L$ can vary based on how the image was acquired. The $z$-resolution is $5$mm in our dataset. Since the venous phase is one of the most informative for diagnosis, and is also ubiquitous in public data, we choose it as the anchor phase and register images from other three phases to it. Consequently, we also synthesize the other three phases images to the venous phase. We divide the dataset into 1350/45/90 patients for training, validation and testing, respectively, and we manaully annotate the liver masks on the validation and testing sets for evaluation. {\it Note that there are in total $1485\times4 = 5940$ 3D CT scans (all containing pathological livers) used in our work. To the best of our knowledge, this is the largest clinically realistic study of this kind to-date.} For the supervised part, we choose a public dataset, i.e., \textbf{MSD} \cite{simpson2019large}, that contains 131 CT images of venous phase with voxel-wise annotations of the liver and divide it into 100/31 for training and validation. We evaluate the performance of all three registration, synthesis and segmentation tasks to measure the impact of joint training.

\subsection{Baseline}
We compare with several strong baselines for all three tasks:
\begin{itemize}
    \item For image synthesis, we choose \textbf{Pix2Pix}~\cite{isola2017image}. We approximately treat the multi-phase CT scans from the same patient as paired data, so that we can better compare to see how incorporating registration can benefit the synthesis module when there is no paired data.
    \item For image registration, we first compare with \textbf{Deeds}~\cite{heinrich2012globally}, one of the best registration methods to date for abdominal CT~\cite{Xu_2016}. The advantage of learning-based methods compared with conventional ones is often on the speed of inference, but we can also show performance improvement. We also compare with the learning-based \textbf{VoxelMorph}~\cite{balakrishnan2019VoxelMorph} with local cross-correlation to handle  multi-modal image registration. 
    \item For the segmentation task, we compare with \textbf{VNet}~\cite{milletari2016v}, which is a popular framework in medical image segmentation.  
\end{itemize}

\begin{table}[tbp]

    \centering
 \caption{Evaluation for the registration task on the CGMH liver dataset in terms
of Dice score, HD (mm), ASD (mm), and GPU/CPU running time (s). Standard deviations are in parentheses.}
    \label{tab:ablation}
    \begin{tabular}{l c c c c c c }
\hline
\multicolumn{1}{l|}{}  & \multicolumn{3}{c|}{Dice $\uparrow$}  
& \multicolumn{3}{c}{HD95 $\downarrow$} \\
\multicolumn{1}{l|}{}  & Arterial & Delay &\multicolumn{1}{c|}{Non-Contrast}
& Arterial & Delay &\multicolumn{1}{c}{Non-Contrast} \\
\hline
\multicolumn{1}{c|}{Initial State} & 90.94 (7.52) &	90.52 (8.08) &	\multicolumn{1}{c|}{90.08 (6.74) } 
& 7.54 (4.89)&	7.86 (5.83)&	\multicolumn{1}{c}{7.87 (4.37)}  \\
\multicolumn{1}{c|}{Affine \cite{marstal2016simpleelastix} } & 92.01 (6.57) &	91.69 (6.80) &	\multicolumn{1}{c|}{91.52 (5.48) } 
& 6.81 (4.83)&	6.95 (5.32)&	\multicolumn{1}{c}{6.73 (3.63)}  \\
\multicolumn{1}{c|}{Deeds \cite{heinrich2012globally}} & 94.73 (2.10)&	94.70 (1.91)&	\multicolumn{1}{c|}{94.73 (1.90)}
& 4.74 (1.96)&	4.76 (1.69)&	\multicolumn{1}{c}{4.62 (1.05)}  \\
\multicolumn{1}{c|}{VoxelMorph \cite{balakrishnan2019VoxelMorph}} & 94.28 (2.53)&	 94.23 (3.15)&	\multicolumn{1}{c|}{93.93 (2.58)}
 &5.29 (2.33)&	5.42 (3.25)&	\multicolumn{1}{c}{5.40 (2.48)} \\
 \hline
\multicolumn{1}{c|}{JSynR-Reg} & 94.81 (2.35)&	94.71 (2.62)&	\multicolumn{1}{c|}{94.57 (2.52)}
 & 4.93 (2.14)&	5.07 (3.06)&		\multicolumn{1}{c}{4.87 (2.30)}\\
\multicolumn{1}{c|}{JSegR-Reg} & 95.52 (1.76)&	95.39 (2.14)&		\multicolumn{1}{c|}{95.37 (1.80)} & 4.47 (2.21)&	4.70 (3.24)&		\multicolumn{1}{c}{4.45 (1.85)}\\
 \multicolumn{1}{c|}{JSSR-Reg} & \textbf{95.56}(1.70)&	 \textbf{95.42}(2.00)&		\multicolumn{1}{c|}{\textbf{95.41}(1.72)}
 & \textbf{4.44}(2.19)&	\textbf{4.65}(3.14)&		\multicolumn{1}{c}{\textbf{4.35}(1.60)}\\

\hline
\multicolumn{1}{l|}{}  & \multicolumn{3}{c|}{ASD $\downarrow$}  
& \multicolumn{3}{c}{Time $\downarrow$} \\
\multicolumn{1}{l|}{}  & Arterial & Delay &\multicolumn{1}{c|}{Non-Contrast}
& Arterial & Delay &\multicolumn{1}{c}{Non-Contrast} \\
\hline
\multicolumn{1}{c|}{Initial State} & 2.12 (1.86) &	2.27 (2.19) &	\multicolumn{1}{c|}{2.37 (1.77) } 
& -/- &	-/-&	\multicolumn{1}{c}{-/-}  \\
\multicolumn{1}{c|}{Affine \cite{marstal2016simpleelastix}} & 1.74 (1.58)&	1.86 (1.89)&	\multicolumn{1}{c|}{1.87 (1.41)} 
& -/7.77&	-/7.77&	\multicolumn{1}{c}{-/7.77}  \\
\multicolumn{1}{c|}{Deeds \cite{heinrich2012globally}} & 1.01 (0.44)&	1.01 (0.39)&	\multicolumn{1}{c|}{0.99 (0.36)} 
& -/41.51&	-/41.51&	\multicolumn{1}{c}{-/41.51}  \\
\multicolumn{1}{c|}{VoxelMorph \cite{balakrishnan2019VoxelMorph}} &1.10 (0.53)&	1.12 (0.87)&	\multicolumn{1}{c|}{1.20 (0.67)} 
 & 1.71/1.76&	1.71/1.76&	\multicolumn{1}{c}{1.71/1.76}\\
 \hline
\multicolumn{1}{c|}{JSynR-Reg} & 0.95 (0.45)&	 0.98 (0.72)&	\multicolumn{1}{c|}{0.98 (0.56)}
 & 3.14/1.76 &	3.14/1.76&		\multicolumn{1}{c}{3.14/1.76}\\

\multicolumn{1}{c|}{JSegR-Reg} & 0.80 (0.37)&	0.83 (0.59) &		\multicolumn{1}{c|}{0.83 (0.40)} 
& 3.14/1.76&	3.14/1.76&		\multicolumn{1}{c}{3.14/1.76} \\
 \multicolumn{1}{c|}{JSSR-Reg} & \textbf{0.79}(0.36)&	\textbf{0.83}(0.56)&		\multicolumn{1}{c|}{\textbf{0.82}(0.37)}
 & 1.71/1.76&	1.71/1.76&		\multicolumn{1}{c}{1.71/1.76}\\

\hline
\end{tabular}
\label{reg_table}
\end{table}

\subsection{Implementation Details}
We conduct several preprocessing procedures. First, since the CT images from different phases, even for the same patient, have different volume sizes, we crop the maximum intersection of all four phases based on the physical coordinates to make their size the same. Second, we apply rigid registration using~\cite{marstal2016simpleelastix} between the four phases, using the venous phase as the anchor. Third, we window the intensity values to $-200$HU to $200$HU and normalize to $-1$ to $1$, and then we resize the CT volume to $256\times 256\times L$ to fit into GPU memory. For the public dataset, we sample along the axial axis to make the resolution also 5mm, and then apply the same intensity preprocessing.

The structure of each component is shown in Fig.~\ref{model}. We choose 3D V-Net~\cite{milletari2016v} for the generator and segmentation module and 3D PHNN~\cite{phnn} for the registration. To optimize the objectives, we use the Adam solver~\cite{kingma2014adam} for all the modules, setting the hyper parameters to $\lambda_{seg}=\lambda_{reg}=1,\ \lambda_{syn}=0.02$. We choose different learning rates for different modules in order to better balance the training:. 0.0001, 0.001, 0.1, and 0.1 for the generator, registration module, segmentation module, and discriminator, respectively. Another way to balance the training is to adjust the loss term weights. However, there are loss terms that relate with multiple modules, which makes it more complex to control each component separately. We train on the Nvidia Quadro RTX 6000 GPU with 24 GB memory, with instance normalization and batch size 1. The training process takes about 1.4 GPU days.

\begin{figure}[!t]
\begin{center}
    \includegraphics[width=\textwidth]{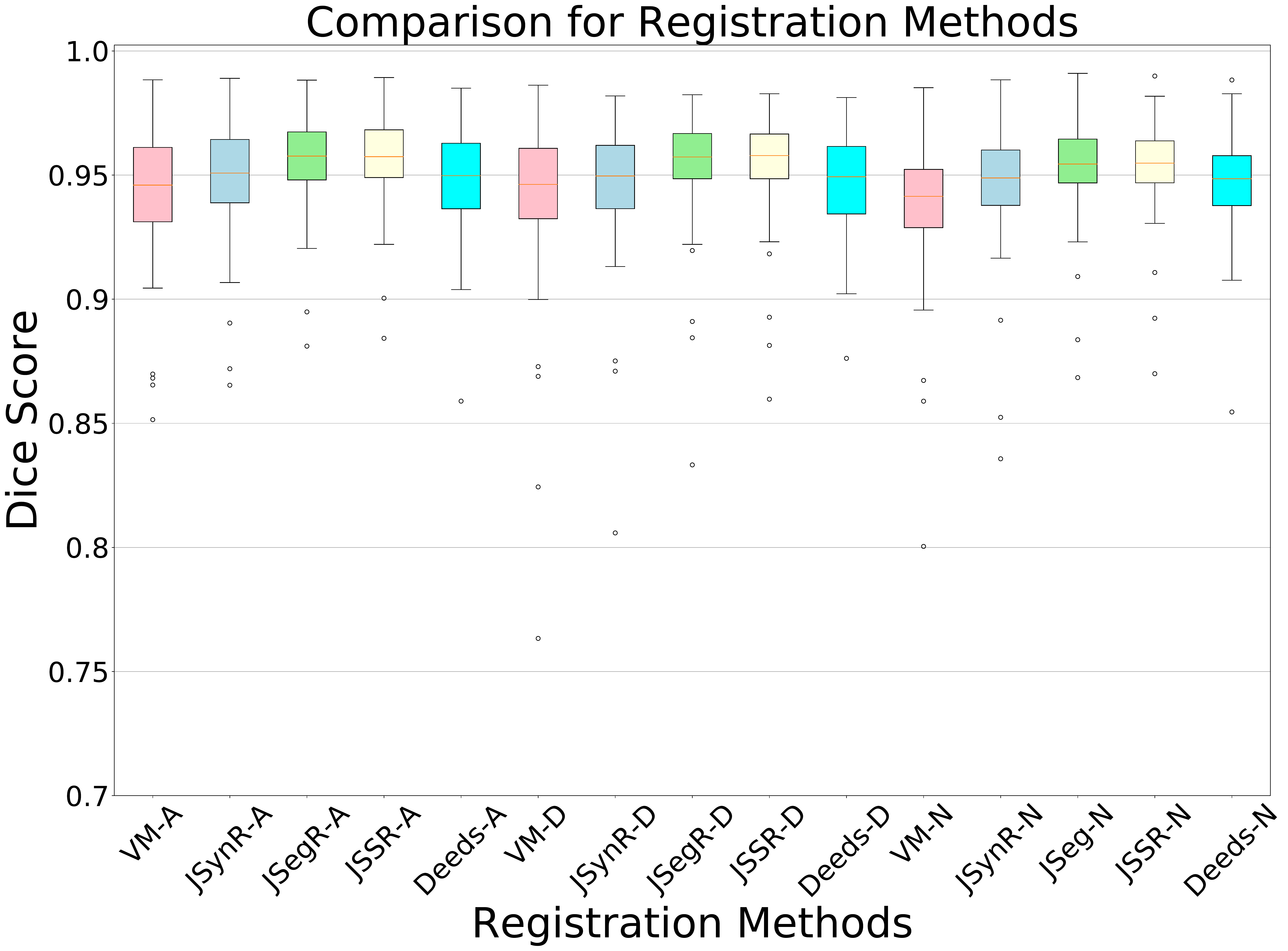}
\end{center}
\caption{
   Box-plots for the registration results (DSC). Suffixes indicate the moving phases (A, D, N for arterial, delay, non-contrast). VM stands for VoxelMorph.
}
\label{Fig2}
\end{figure}

\subsection{Main Results}
\subsubsection{Multi-modal image registration} 
We summarize the results of registration task in Table~\ref{reg_table}. We use the manual annotations of the test set and evaluate the similarity between those of fixed image, which is always in the venous phase here, and the warped labels of the moving images chosen from arterial, delay and non-contrast.  The similarity is measured using the Dice score, 95 percent hausdorff distance (HD), and the average surface distance (ASD). We also report the consumed time on GPU/CPU in sec for each method. We use the term ``Initial State'' to refer to the result before applying any registration and ``Affine'' to the result after rigid registration. We denote our joint system as JSSR and JSSR-Reg is only the registration part of JSSR. We also compare two ablations of JSSR. JSynR, which only contains the generator and registration module, is optimized using \eqref{bi}. JSegR has the segmentation and registration module instead. More details will be discussed in Section $5$. As can be seen, our JSSR method outperforms Deeds by $0.83\%$ by average Dice, while executing much faster in terms of inference. Also by taking advantage of the joint training, JSSR achieves significantly higher results than VoxelMorph (exceeded by $1.28\%$) with comparable inference time. We can observe gradual improvements from VoxelMorph to JSynR to JSSR, which demonstrates the successive contributions of joint training. Fig.~\ref{Fig2} depicts a box plot of these results. 

\begin{table}[tbp]
    \centering
 \caption{Evaluation for the synthesis and segmentation tasks on the CGMH liver dataset in terms
of average Dice score}

    \begin{tabular}{l c c c c }
\hline
\multicolumn{1}{l|}{} 
& \multicolumn{4}{c}{VNet \cite{milletari2016v}} \\
\multicolumn{1}{c|}{Dice $\uparrow$}  & Venous & Arterial & Delay & \multicolumn{1}{c}{Non-Contrast}  \\
\hline
\multicolumn{1}{l|}{No-Synthesis} & 90.47 (6.23)&	89.47 (7.05)&	89.88 (6.38)&	\multicolumn{1}{c}{89.38 (6.38)}  \\
\multicolumn{1}{l|}{Pix2Pix \cite{isola2017image} } & 90.47 (6.23)&	76.50 (17.77)&	79.60 (13.13)&	\multicolumn{1}{c}{67.48 (15.97)}\\
 \hline
\multicolumn{1}{l|}{JSynR-Syn} & 90.47 (6.23)&	89.69 (7.09)&	90.01 (6.27) &	\multicolumn{1}{c}{90.15 (6.21)}\\
\multicolumn{1}{l|}{JSSR-Syn} & 90.47 (6.23)&	89.44 (7.15)&	89.76 (6.34)&	\multicolumn{1}{c}{89.31 (7.57)} \\
\hline
\hline
\multicolumn{1}{l|}{} 
& \multicolumn{4}{c}{JSegR-Seg} \\
\multicolumn{1}{c|}{Dice $\uparrow$}  & Venous & Arterial & Delay & \multicolumn{1}{c}{Non-Contrast}  \\
\hline
\multicolumn{1}{l|}{No-Synthesis} & 91.88 (4.84)&	90.91 (5.06)&	91.18 (4.68) &	\multicolumn{1}{c}{91.12 (4.72)}  \\
\multicolumn{1}{l|}{Pix2Pix \cite{isola2017image}} & 91.88 (4.84)&	89.59 (5.51)&	87.78 (5.78)&	\multicolumn{1}{c}{89.59 (5.51)}\\
 \hline
\multicolumn{1}{l|}{JSynR-Syn} & 91.88 (4.84)&	91.15 (4.93)&	91.37 (4.56)&	\multicolumn{1}{c}{91.36 (4.54)}\\
\multicolumn{1}{l|}{JSSR-Syn} & 91.88 (4.84)&	91.12 (4.99)&	91.30 (4.63)&	\multicolumn{1}{c}{91.39 (4.53)} \\
\hline
\hline
\multicolumn{1}{l|}{} 
& \multicolumn{4}{c}{JSSR-Seg} \\
\multicolumn{1}{c|}{Dice $\uparrow$}  & Venous & Arterial & Delay & \multicolumn{1}{c}{Non-Contrast}  \\
\hline
\multicolumn{1}{l|}{No-Synthesis} & 92.24 (3.88)&	91.25 (4.10)&	91.34 (3.76)&	\multicolumn{1}{c}{91.37 (3.81)}\\
\multicolumn{1}{l|}{Pix2Pix \cite{isola2017image}} & 92.24 (3.88) &	85.30 (7.11) &	84.68 (9.29)&	\multicolumn{1}{c}{79.89 (8.49)}\\
 \hline
\multicolumn{1}{l|}{JSynR-Syn} & 92.24 (3.88)&	91.42 (4.06)&	91.58 (3.64)&	\multicolumn{1}{c}{91.67 (3.67)}\\
\multicolumn{1}{l|}{JSSR-Syn} & 92.24 (3.88)&	91.39 (4.10)&	91.51 (3.72)&	\multicolumn{1}{c}{91.60 (3.69)}\\
\hline
\end{tabular}
\label{synseg} 
\end{table}


\subsubsection{Multi-modal image segmentation and synthesis}
\begin{figure}[tbp]
  \centering
\begin{subfigure}{.7\textwidth}

  \centering
  \includegraphics[width=1\linewidth]{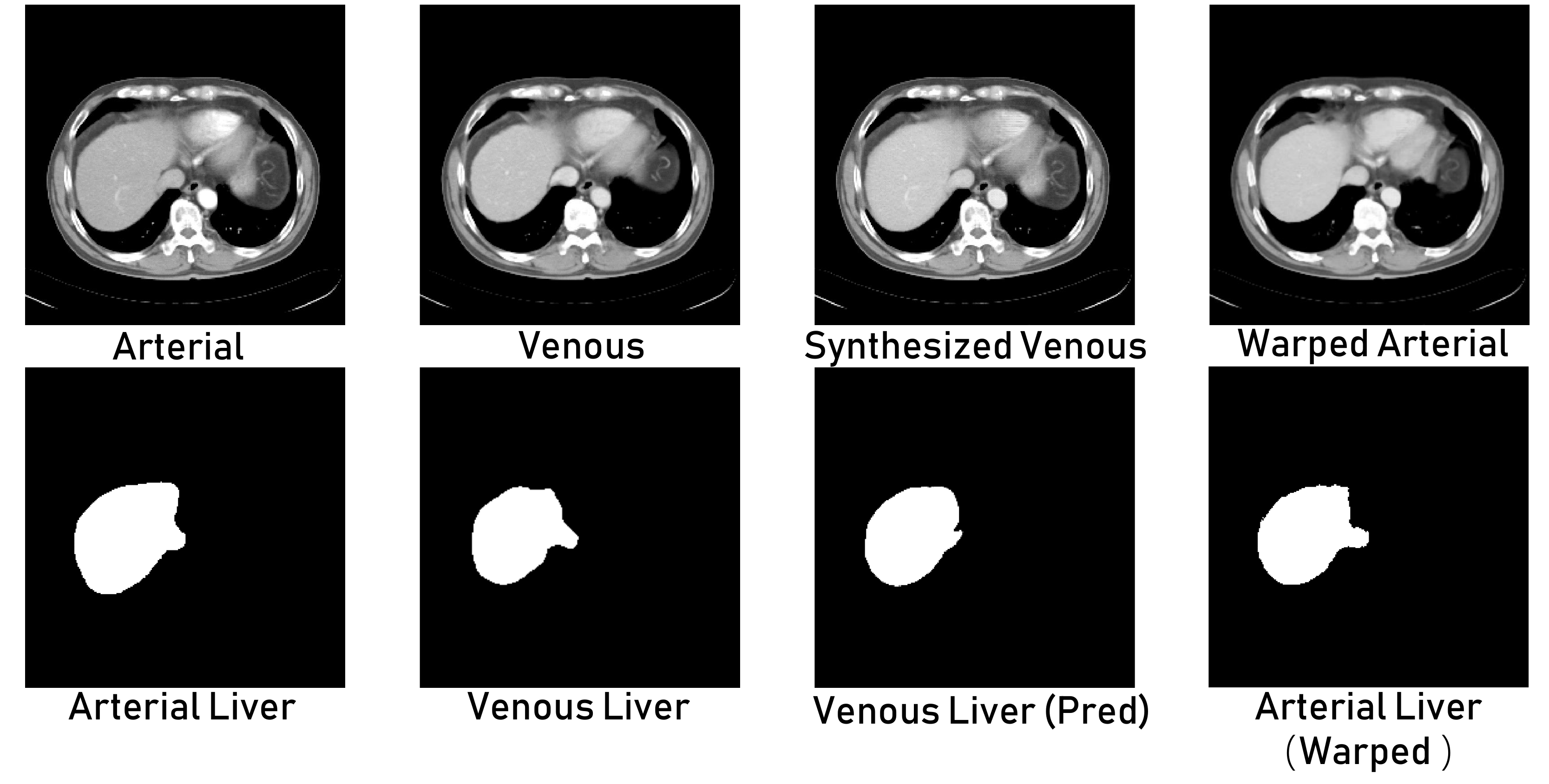}  
  \caption{Results on the arterial CT phase.}
  \label{fig:sub-first} 
\end{subfigure}

\begin{subfigure}{.7\textwidth}
  \centering
  \includegraphics[width=1\linewidth]{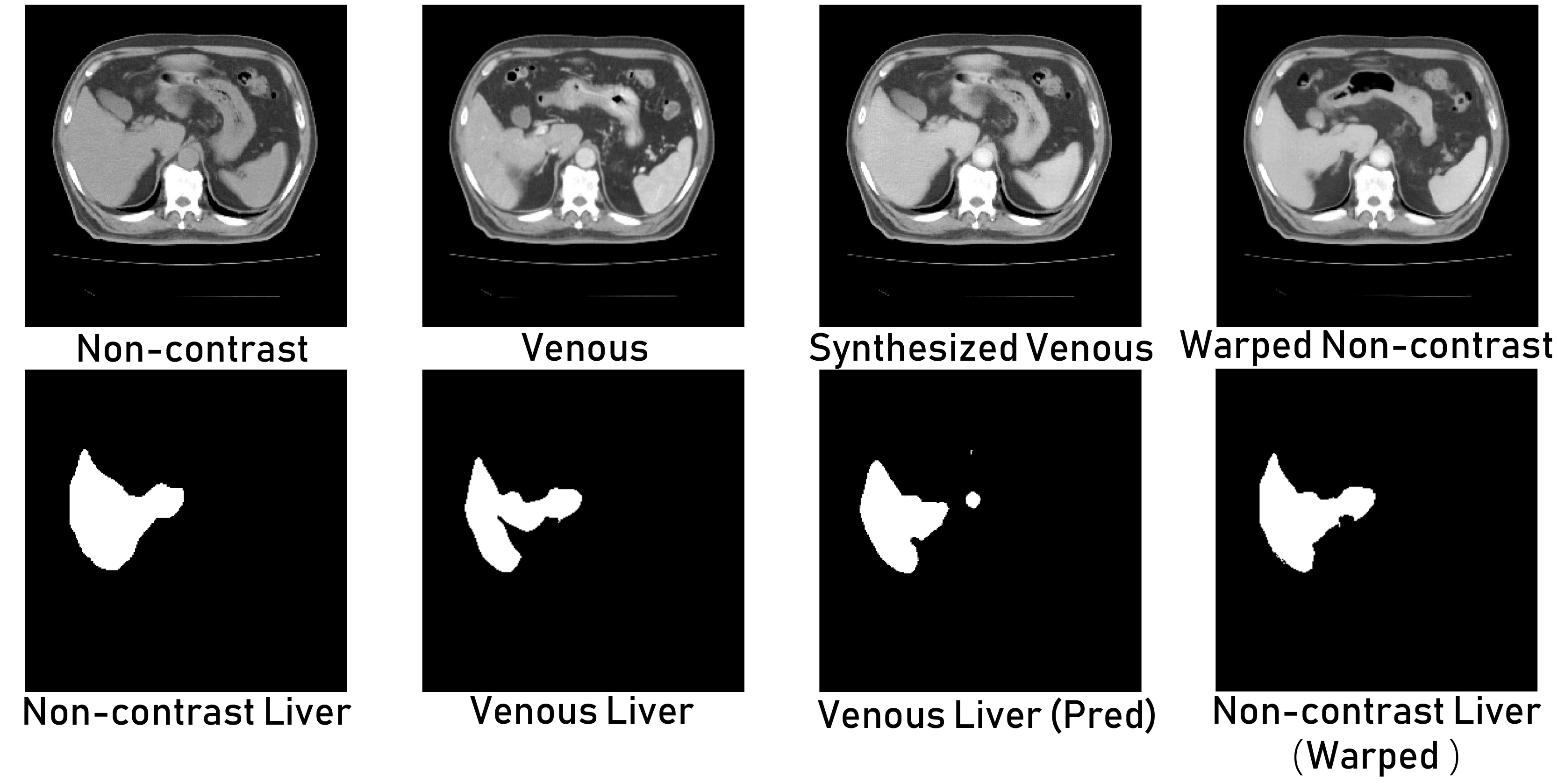}  
  \caption{Results on the non-contrast CT phase.}
  \label{fig:sub-third} 
\end{subfigure}

\caption{Qualitative examples of JSSR synthesis, segmentation and registration.}
\label{Fig3}
\end{figure}

Table~\ref{synseg} presents the synthesis and segmentation evaluations. Following the practice of \cite{isola2017image}, we evaluate the synthesis model by applying the segmentation model on the synthesized image. The intuition is that the better the synthesized image is, the better the segmentation map can be estimated. We evaluate with three segmentation models. The VNet baseline is trained on the MSD dataset with full supervision. JSegR-Seg is the segmentation part of JSegR as described in Section $5$. JSSR-Seg is the segmentation module of our JSSR system. For each segmentation model, we test it on different synthesis model, thus comparing all possible synthesis/segmentation combinations. For ``No-Synthesis'', we directly apply the segmentation model on original images. For the  three synthesis models, we test the segmentation model on the original venous image and also on the ``fake'' venous images synthesized from arterial, delay, non-contrast phases. From the No-Synthesis lines we can observe a clear performance drop when directly applying the segmentation model to arterial, delay and non-contrast phases, since the supervised data is all from the venous phase.  For Pix2Pix, the performance goes through different levels of reduction among different segmentation algorithms and is not as high as the Non-Synthesis. That may be caused by artifacts introduced by the GAN model and the L1 term is providing less constraint since there is no paired data. Comparing the JSynR-Syn and JSSR-Syn generators, the performance is improved by creating true paired data via the registration process, but even so, it is just comparable to No-Synthesis. For JSynR-Syn, the JSynR is not jointly learned with a segmentation process, so the performance for synthesized images does not necessarily go up. For JSSR-Syn, however, it means the constraints we are using for optimizing the system does not bring enough communication between the generator and segmentor to improve the former. Even so, we can improvements from VNet to JSegR-Seg to JSSR-Seg on both the No-Synthesis and various synthesis options, indicating that the segmentation process can still benefit from a joint system, which includes the synthesis module. Please refer to Fig.~\ref{Fig3} for qualitative examples of JSSR registration, synthesis and segmentation results.

\section{Ablation and Discussion}
\subsubsection{JSegR vs JSSR}
We implement JSegR as another ablation. The purpose is to explore the importance of the synthesis module for the JSSR system. Since JSegR does not have a generator, the registration module takes images from different phases directly as input. The segmentation consistency term in (\ref{segunsup}) is then replaced with
\begin{equation}\label{vseg}
    \mathcal{L}_{dice}^{reg}(S,\Phi)=E_{x,y}1-Dice[\tau(S(x)),S(y)],
\end{equation}
where $\tau=\Phi(x,y)$. This framework is similar to \cite{xu2019deepatlas}, which jointly learned the registration and the segmentation module . In our case, though, $x,y$ are in a different domain and the annotations are unavailable. This method is expected to struggle, since $x,y$ are in different phases. However, as shown in Table~\ref{synseg}, the performance drop across phases is not too severe even for the baseline VNet. Correspondingly, JSegR can achieve a higher result on registration than JSynR and performs close to JSSR, which demonstrates the great importance of incorporating semantic information into the registration. 

\subsubsection{Extra constraints}

The constraints detailed in Fig.~\ref{motivation} are not the only possible constraints. For instance, constraints can be added to ensure consistency between ``register first'' vs ``register last'' pipelines:
\begin{equation}
    \mathcal{L}_{L1}^{reg}(\Phi,G)=E_{x,y}||G(\tau(x),y)-\tau(G(x,y))||_1. \label{eqn:extra}
\end{equation}
However, each constraint introduces additional complexity. Future work should explore whether \eqref{eqn:extra}, or other constraints, can boost performance further. 



\section{Conclusion}
In this paper, we propose a novel JSSR system for multi-modal image registration. Our system takes advantages of joint learning based on the intrinsic  connections between the synthesis, segmentation and registration tasks. The optimization can be conducted end-to-end with several unsupervised consistency loss and each component benefits from the joint training process. We evaluate the JSSR system on a large-scale multi-phase clinically realistic CT image dataset without any segmentation annotations. After joint training, the performance of registration and segmentation increases by $0.91\%$ and $1.86\%$ respectively on the average Dice score for all the phases. Our system outperforms the recent VoxelMorph algorithm~\cite{balakrishnan2019VoxelMorph} by $1.28\%$, and the state-of-the-art conventional multi-modal registration method~\cite{heinrich2012globally} by $0.83\%$, but has considerably faster inference time. 

\noindent
{\bf Acknowledgements.}
This work was partially supported by the Lustgarten Foundation for Pancreatic Cancer Research. The main work was done when F. Liu was a research Intern at PAII Inc. We thank Zhuotun Zhu and Yingda Xia for instructive discussions.

\clearpage
%
%
\bibliographystyle{splncs04}
\bibliography{egbib}

\begin{thebibliography}{10}
\providecommand{\url}[1]{\texttt{#1}}
\providecommand{\urlprefix}{URL }
\providecommand{\doi}[1]{https://doi.org/#1}

\bibitem{arar2020unsupervised}
Arar, M., Ginger, Y., Danon, D., Bermano, A.H., Cohen-Or, D.: Unsupervised
  multi-modal image registration via geometry preserving image-to-image
  translation. In: Proceedings of the IEEE/CVF Conference on Computer Vision
  and Pattern Recognition. pp. 13410--13419 (2020)

\bibitem{avants2008symmetric}
Avants, B.B., Epstein, C.L., Grossman, M., Gee, J.C.: Symmetric diffeomorphic
  image registration with cross-correlation: evaluating automated labeling of
  elderly and neurodegenerative brain. Medical image analysis  \textbf{12}(1),
  26--41 (2008)

\bibitem{balakrishnan2019VoxelMorph}
Balakrishnan, G., Zhao, A., Sabuncu, M.R., Guttag, J., Dalca, A.V.: Voxelmorph:
  a learning framework for deformable medical image registration. IEEE
  transactions on medical imaging  \textbf{38}(8),  1788--1800 (2019)

\bibitem{blendowski2019learning}
Blendowski, M., Heinrich, M.P.: Learning interpretable multi-modal features for
  alignment with supervised iterative descent. In: International Conference on
  Medical Imaging with Deep Learning. pp. 73--83 (2019)

\bibitem{cao2017dual}
Cao, X., Yang, J., Gao, Y., Guo, Y., Wu, G., Shen, D.: Dual-core steered
  non-rigid registration for multi-modal images via bi-directional image
  synthesis. Medical image analysis  \textbf{41},  18--31 (2017)

\bibitem{phnn}
Harrison, A.P., Xu, Z., George, K., Lu, L., Summers, R.M., Mollura, D.J.:
  Progressive and multi-path holistically nested neural networks for
  pathological lung segmentation from ct images. In: International Conference
  on Medical Image Computing and Computer-Assisted Intervention. pp. 621--629.
  Springer (2017)

\bibitem{heinrich2015multi}
Heinrich, M., Maier, O., Handels, H.: Multi-modal multi-atlas segmentation
  using discrete optimisation and self-similarities. CEUR Workshop Proceedings
  (2015)

\bibitem{heinrich2012mind}
Heinrich, M.P., Jenkinson, M., Bhushan, M., Matin, T., Gleeson, F.V., Brady,
  M., Schnabel, J.A.: Mind: Modality independent neighbourhood descriptor for
  multi-modal deformable registration. Medical image analysis  \textbf{16}(7),
  1423--1435 (2012)

\bibitem{heinrich2012globally}
Heinrich, M.P., Jenkinson, M., Brady, M., Schnabel, J.A.: Globally optimal
  deformable registration on a minimum spanning tree using dense displacement
  sampling. In: International Conference on Medical Image Computing and
  Computer-Assisted Intervention. pp. 115--122. Springer (2012)

\bibitem{heinrich2013mrf}
Heinrich, M.P., Jenkinson, M., Brady, M., Schnabel, J.A.: Mrf-based deformable
  registration and ventilation estimation of lung ct. IEEE transactions on
  medical imaging  \textbf{32}(7),  1239--1248 (2013)

\bibitem{heinrich2013towards}
Heinrich, M.P., Jenkinson, M., Papie{\.z}, B.W., Brady, M., Schnabel, J.A.:
  Towards realtime multimodal fusion for image-guided interventions using
  self-similarities. In: International conference on medical image computing
  and computer-assisted intervention. pp. 187--194. Springer (2013)

\bibitem{hu2018weakly}
Hu, Y., Modat, M., Gibson, E., Li, W., Ghavami, N., Bonmati, E., Wang, G.,
  Bandula, S., Moore, C.M., Emberton, M., et~al.: Weakly-supervised
  convolutional neural networks for multimodal image registration. Medical
  image analysis  (2018)

\bibitem{huang2018multimodal}
Huang, X., Liu, M.Y., Belongie, S., Kautz, J.: Multimodal unsupervised
  image-to-image translation. In: Proceedings of the European Conference on
  Computer Vision (ECCV). pp. 172--189 (2018)

\bibitem{huo2018synseg}
Huo, Y., Xu, Z., Moon, H., Bao, S., Assad, A., Moyo, T.K., Savona, M.R.,
  Abramson, R.G., Landman, B.A.: Synseg-net: Synthetic segmentation without
  target modality ground truth. IEEE transactions on medical imaging  (2018)

\bibitem{isola2017image}
Isola, P., Zhu, J.Y., Zhou, T., Efros, A.A.: Image-to-image translation with
  conditional adversarial networks. In: Proceedings of the IEEE conference on
  computer vision and pattern recognition. pp. 1125--1134 (2017)

\bibitem{jaderberg2015spatial}
Jaderberg, M., Simonyan, K., Zisserman, A., et~al.: Spatial transformer
  networks. In: Advances in neural information processing systems. pp.
  2017--2025 (2015)

\bibitem{ketcha2019learning}
Ketcha, M.D., De~Silva, T.S., Han, R., Uneri, A., Vogt, S., Kleinszig, G.,
  Siewerdsen, J.H.: Learning-based deformable image registration: effect of
  statistical mismatch between train and test images. Journal of Medical
  Imaging  (2019)

\bibitem{kingma2014adam}
Kingma, D.P., Ba, J.: Adam: {A} method for stochastic optimization. In: 3rd
  International Conference on Learning Representations, {ICLR} 2015, San Diego,
  CA, USA, May 7-9, 2015, Conference Track Proceedings (2015)

\bibitem{li2019hybrid}
Li, B., Niessen, W.J., Klein, S., de~Groot, M., Ikram, M.A., Vernooij, M.W.,
  Bron, E.E.: A hybrid deep learning framework for integrated segmentation and
  registration: evaluation on longitudinal white matter tract changes. In:
  International Conference on Medical Image Computing and Computer-Assisted
  Intervention. pp. 645--653. Springer (2019)

\bibitem{li2017non}
Li, H., Fan, Y.: Non-rigid image registration using fully convolutional
  networks with deep self-supervision. arXiv preprint arXiv:1709.00799  (2017)

\bibitem{maes1997multimodality}
Maes, F., Collignon, A., Vandermeulen, D., Marchal, G., Suetens, P.:
  Multimodality image registration by maximization of mutual information. IEEE
  transactions on Medical Imaging  \textbf{16}(2),  187--198 (1997)

\bibitem{mahapatra2018deformable}
Mahapatra, D., Antony, B., Sedai, S., Garnavi, R.: Deformable medical image
  registration using generative adversarial networks. In: 2018 IEEE 15th
  International Symposium on Biomedical Imaging (ISBI 2018). pp. 1449--1453.
  IEEE (2018)

\bibitem{maintz1996overview}
Maintz, J.A., Viergever, M.A.: An overview of medical image registration
  methods. In: Symposium of the Belgian hospital physicists association
  (SBPH/BVZF). vol.~12, pp. 1--22. Citeseer (1996)

\bibitem{marstal2016simpleelastix}
Marstal, K., Berendsen, F., Staring, M., Klein, S.: Simpleelastix: A
  user-friendly, multi-lingual library for medical image registration. In:
  Proceedings of the IEEE conference on computer vision and pattern recognition
  workshops (2016)

\bibitem{milletari2016v}
Milletari, F., Navab, N., Ahmadi, S.A.: V-net: Fully convolutional neural
  networks for volumetric medical image segmentation. In: 2016 Fourth
  International Conference on 3D Vision (3DV). pp. 565--571. IEEE (2016)

\bibitem{qin2018joint}
Qin, C., Bai, W., Schlemper, J., Petersen, S.E., Piechnik, S.K., Neubauer, S.,
  Rueckert, D.: Joint learning of motion estimation and segmentation for
  cardiac mr image sequences. In: International Conference on Medical Image
  Computing and Computer-Assisted Intervention. pp. 472--480. Springer (2018)

\bibitem{qin2019unsupervised}
Qin, C., Shi, B., Liao, R., Mansi, T., Rueckert, D., Kamen, A.: Unsupervised
  deformable registration for multi-modal images via disentangled
  representations. In: International Conference on Information Processing in
  Medical Imaging. pp. 249--261. Springer (2019)

\bibitem{landmark2001}
{Rohr}, K., {Stiehl}, H.S., {Sprengel}, R., {Buzug}, T.M., {Weese}, J., {Kuhn},
  M.H.: Landmark-based elastic registration using approximating thin-plate
  splines. IEEE Transactions on Medical Imaging  (2001)

\bibitem{shechtman2007matching}
Shechtman, E., Irani, M.: Matching local self-similarities across images and
  videos. In: 2007 IEEE Conference on Computer Vision and Pattern Recognition
  (2007)

\bibitem{simpson2019large}
Simpson, A.L., Antonelli, M., Bakas, S., Bilello, M., Farahani, K.,
  Van~Ginneken, B., Kopp-Schneider, A., Landman, B.A., Litjens, G., Menze, B.,
  et~al.: A large annotated medical image dataset for the development and
  evaluation of segmentation algorithms. arXiv preprint arXiv:1902.09063
  (2019)

\bibitem{studholme1999overlap}
Studholme, C., Hill, D.L., Hawkes, D.J.: An overlap invariant entropy measure
  of 3d medical image alignment. Pattern recognition  \textbf{32}(1),  71--86
  (1999)

\bibitem{landmark2019}
Sultana, S., Song, D.Y., Lee, J.: {A deformable multimodal image registration
  using PET/CT and TRUS for intraoperative focal prostate brachytherapy}. In:
  Fei, B., Linte, C.A. (eds.) Medical Imaging 2019: Image-Guided Procedures,
  Robotic Interventions, and Modeling. vol. 10951, pp. 383 -- 388.
  International Society for Optics and Photonics, SPIE (2019).
  \doi{10.1117/12.2512996}

\bibitem{tanner2018generative}
Tanner, C., Ozdemir, F., Profanter, R., Vishnevsky, V., Konukoglu, E., Goksel,
  O.: Generative adversarial networks for mr-ct deformable image registration.
  arXiv preprint arXiv:1807.07349  (2018)

\bibitem{de2019deep}
de~Vos, B.D., Berendsen, F.F., Viergever, M.A., Sokooti, H., Staring, M.,
  I{\v{s}}gum, I.: A deep learning framework for unsupervised affine and
  deformable image registration. Medical image analysis  \textbf{52},  128--143
  (2019)

\bibitem{de2017end}
de~Vos, B.D., Berendsen, F.F., Viergever, M.A., Staring, M., I{\v{s}}gum, I.:
  End-to-end unsupervised deformable image registration with a convolutional
  neural network. In: Deep Learning in Medical Image Analysis and Multimodal
  Learning for Clinical Decision Support, pp. 204--212. Springer (2017)

\bibitem{wei2019synthesis}
Wei, D., Ahmad, S., Huo, J., Peng, W., Ge, Y., Xue, Z., Yap, P.T., Li, W.,
  Shen, D., Wang, Q.: Synthesis and inpainting-based mr-ct registration for
  image-guided thermal ablation of liver tumors. In: International Conference
  on Medical Image Computing and Computer-Assisted Intervention. pp. 512--520.
  Springer (2019)

\bibitem{AOODS2009495}
Woods, R.P.: Handbook of medical image processing and analysis (2009)

\bibitem{Xu_2016}
{Xu}, Z., {Lee}, C.P., {Heinrich}, M.P., {Modat}, M., {Rueckert}, D.,
  {Ourselin}, S., {Abramson}, R.G., {Landman}, B.A.: Evaluation of six
  registration methods for the human abdomen on clinically acquired ct. IEEE
  Transactions on Biomedical Engineering  \textbf{63}(8),  1563--1572 (2016)

\bibitem{xu2019deepatlas}
Xu, Z., Niethammer, M.: Deepatlas: Joint semi-supervised learning of image
  registration and segmentation. In: International Conference on Medical Image
  Computing and Computer-Assisted Intervention. pp. 420--429. Springer (2019)

\bibitem{Yang2011}
Yang, X., Akbari, H., Halig, L., Fei, B.: 3d non-rigid registration using
  surface and local salient features for transrectal ultrasound image-guided
  prostate biopsy. Proceedings of SPIE--the International Society for Optical
  Engineering  \textbf{7964},  79642V--79642V (2011). \doi{10.1117/12.878153}

\bibitem{zhang2018translating}
Zhang, Z., Yang, L., Zheng, Y.: Translating and segmenting multimodal medical
  volumes with cycle-and shape-consistency generative adversarial network. In:
  Proceedings of the IEEE conference on computer vision and pattern Recognition
  (2018)

\bibitem{Zheng2018PhaseCN}
Zheng, H., Xie, L., Ni, T., Zhang, Y., Wang, Y.F., Tian, Q., Fishman, E.K.,
  Yuille, A.L.: Phase collaborative network for multi-phase medical imaging
  segmentation. ArXiv  \textbf{abs/1811.11814} (2018)

\bibitem{zhou2019hyper}
Zhou, Y., Li, Y., Zhang, Z., Wang, Y., Wang, A., Fishman, E.K., Yuille, A.L.,
  Park, S.: Hyper-pairing network for multi-phase pancreatic ductal
  adenocarcinoma segmentation. In: International Conference on Medical Image
  Computing and Computer-Assisted Intervention. pp. 155--163. Springer (2019)

\end{thebibliography}
\end{document}